\documentstyle[12pt]{article}
\newlength{\extraspace}
\newlength{\extraspaces}
\newcommand{\be}{\begin{equation}
\addtolength{\abovedisplayskip}{\extraspaces}
\addtolength{\belowdisplayskip}{\extraspaces}
\addtolength{\abovedisplayshortskip}{\extraspace}
\addtolength{\belowdisplayshortskip}{\extraspace}}
\newcommand{\ee}{\end{equation}}

\newcommand{\ba}{\begin{eqnarray}
\addtolength{\abovedisplayskip}{\extraspaces}
\addtolength{\belowdisplayskip}{\extraspaces}
\addtolength{\abovedisplayshortskip}{\extraspace}
\addtolength{\belowdisplayshortskip}{\extraspace}}
\newcommand{\ea}{\end{eqnarray}}

\newcommand{\newsection}[1]{
\vspace{15mm}
\pagebreak[3]
\addtocounter{section}{1}
\setcounter{equation}{0}
\setcounter{subsection}{0}
\setcounter{footnote}{0}
\begin{flushleft}
{\large\bf \thesection. #1}
\end{flushleft}
\nopagebreak
\medskip
\nopagebreak}

\newcommand{\Tr}{{\rm Tr}}

\begin{document}

\addtolength{\baselineskip}{.8mm}

{\thispagestyle{empty}
\noindent \hspace{1cm}  \hfill IFUP--TH/2001--30 \hspace{1cm}\\
\mbox{}                 \hfill October 2001 \hspace{1cm}\\

\begin{center}
\vspace*{1.0cm}
{\large\bf The analytic continuation of the high--energy} \\
{\large\bf parton--parton scattering amplitude with an IR cutoff} \\
\vspace*{1.0cm}
{\large Enrico Meggiolaro\footnote{E--mail:
enrico.meggiolaro@df.unipi.it} }\\
\vspace*{0.5cm}{\normalsize
{Dipartimento di Fisica, \\
Universit\`a di Pisa, \\
Via Buonarroti 2, \\
I--56127 Pisa, Italy.}}\\
\vspace*{2cm}{\large \bf Abstract}
\end{center}

\noindent
The high--energy parton--parton scattering amplitude can be described, in the
c.m.s.,  by the expectation value of two infinite Wilson lines, running
along the classical trajectories of the two colliding particles.
The above description suffers from IR divergences (typical of $3 + 1$
dimensional gauge theories), which can be regularized by considering
finite Wilson lines, extending in proper time from $-T$ to $T$ (and
eventually letting $T \to +\infty$).
Generalizing the results of a previous paper, we give here the general proof
that the expectation value of two IR--regularized  Wilson lines, forming a
certain hyperbolic angle in Minkowski space--time, and the expectation value
of two IR--regularized Euclidean Wilson lines, forming a certain angle in
Euclidean four--space, are connected by an analytic continuation in the angular
variables and in the IR cutoff $T$. This result can be used to evaluate the
IR--regularized high--energy scattering amplitude directly in the Euclidean
theory. \\
}
\vfill\eject

\newsection{Introduction}

\noindent
The parton--parton scattering amplitude, at high squared
energies $s$ in the center of mass and small squared transferred momentum $t$
(that is $s \to \infty$ and $|t| \ll s$, let us say $|t| \le 1~{\rm GeV}^2$),
can be described by the expectation value of two infinite Wilson lines,
running along the classical trajectories of the two colliding particles
\cite{Nachtmann91,Nachtmann97,Meggiolaro96,Meggiolaro01}.

Let us consider, for example, the case of the quark--quark scattering
amplitude.
If one defines the scattering amplitude $T_{fi} = \langle f | \hat{T} |
i \rangle$, between the initial state $| i \rangle$ and the final
state $| f \rangle$, as follows ($\hat{S}$ being the scattering operator)
\be
\langle f | ( \hat{S} - {\bf 1} ) | i \rangle
= i (2\pi)^4 \delta^{(4)} (P_{fin} - P_{in})
~\langle f | \hat{T} | i \rangle ~,
\label{Tfi}
\ee
where $P_{in}$ is the initial total four--momentum and $P_{fin}$ is the final
total four--momentum, then, in the center--of--mass reference system (c.m.s.),
taking for example the initial trajectories of the two quarks along the
$x^1$--axis, the high--energy scattering amplitude $T_{fi}$ has the following
form [explicitly indicating the color indices ($i,j, \ldots$)
and the spin indices ($\alpha, \beta, \ldots$) of the quarks]
\cite{Nachtmann91,Nachtmann97,Meggiolaro96,Meggiolaro01}
\ba
\lefteqn{
T_{fi} = \langle \psi_{i\alpha}(p'_1) \psi_{k\gamma}(p'_2) | \hat{T} |
\psi_{j\beta}(p_1) \psi_{l\delta}(p_2) \rangle } \nonumber \\
& & \mathop{\sim}_{s \to \infty}
-{i \over Z_W^2} \cdot \delta_{\alpha\beta} \delta_{\gamma\delta}
\cdot 2s
\displaystyle\int d^2 \vec{z}_\perp e^{i \vec{q}_\perp \cdot \vec{z}_\perp}
\langle [ W_{p_1} (z_t) - {\bf 1} ]_{ij} [ W_{p_2} (0) - {\bf 1} ]_{kl} \rangle
~,
\label{scatt1}
\ea
where $q = (0,0,\vec{q}_\perp)$, with $t = q^2 = -\vec{q}_\perp^2$, is the
tranferred four--momentum and $z_t = (0,0,\vec{z}_\perp)$, with $\vec{z}_\perp
= (z^2,z^3)$, is the distance between the two trajectories in the
{\it transverse} plane
[the coordinates $(x^0,x^1)$ are often called {\it longitudinal} coordinates].
The expectation
value $\langle f(A) \rangle$ is the average of $f(A)$ in the sense of the
functional integration over the gluon field $A^\mu$ (including also the
determinant of the fermion matrix, i.e., $\det[i\gamma^\mu D_\mu - m_0]$,
where $D^\mu = \partial^\mu + ig A^\mu$ is the covariant derivative and $m_0$
is the {\it bare} quark mass).
The two infinite Wilson lines $W_{p_1} (z_t)$ and $W_{p_2} (0)$ in
Eq. (\ref{scatt1}) are defined as
\ba
W_{p_1} (z_t) &=&
{\cal T} \exp \left[ -ig \displaystyle\int_{-\infty}^{+\infty}
A_\mu (z_t + p_1 \tilde{\tau}) p_1^\mu d\tilde{\tau} \right] ~;
\nonumber \\
W_{p_2} (0) &=&
{\cal T} \exp \left[ -ig \displaystyle\int_{-\infty}^{+\infty}
A_\mu (p_2 \tilde{\tau}) p_2^\mu d\tilde{\tau} \right] ~,
\label{Wilson0}
\ea
where ${\cal T}$ stands for ``{\it time ordering}'' and $A_\mu = A_\mu^a T^a$;
the four--vectors $p_1 \simeq (E,E,0,0)$ and $p_2 \simeq (E,-E,0,0)$ are the
initial four--momenta of the two quarks [$s = (p_1 + p_2)^2 = 4E^2$].
The space--time configuration of these two Wilson lines is shown in Fig. 1.

\newcommand{\wilson}[1]{
\begin{figure}
\begin{center}
\setlength{\unitlength}{1.00mm}
\raisebox{-40\unitlength}
{\mbox{\begin{picture}(80,45)(-35,-30)
\thicklines
\put(-22,22){\line(1,-1){41}}
\put(-15,15){\vector(-1,1){1}}
\put(16,23){\line(-1,-1){42}}
\put(9,16){\vector(1,1){1}}
\put(0,0){\vector(-2,-1){13}}
\thinlines
\put(-8,0){\line(1,0){35}}
\put(8,4){\line(-2,-1){35}}
\put(0,-8){\line(0,1){35}}
\put(27,0){\vector(1,0){1}}
\put(0,27){\vector(0,1){1}}
\put(-13,20){\makebox(0,0){$W_{p_2}$}}
\put(20,20){\makebox(0,0){$W_{p_1}$}}
\put(-6,-7){\makebox(0,0){$z_t$}}
\put(25,-3){\makebox(0,0){$x^1$}}
\put(-3,25){\makebox(0,0){$x^0$}}
\end{picture}}}
\parbox{13cm}{\small #1}
\end{center}
\end{figure}}

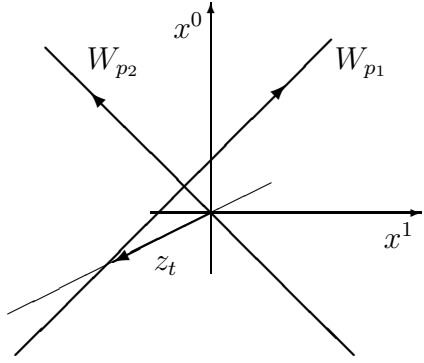
\wilson{{\bf Fig.~1.} The space--time configuration of the two Wilson
lines $W_{p_1}$ and $W_{p_2}$ entering in the expression (\ref{scatt1}) for
the quark--quark elastic scattering amplitude in the high--energy limit.}

Finally, $Z_W$ in Eq. (\ref{scatt1}) is the residue at the pole (i.e., for
$p^2 \to m^2$, $m$ being the quark {\it pole} mass) of the unrenormalized
quark propagator, which can be written in the eikonal approximation as
\cite{Nachtmann91,Meggiolaro01}
\be
Z_W \simeq {1 \over N_c} \langle \Tr [ W_{p_1} (z_t) ] \rangle
= {1 \over N_c} \langle \Tr [ W_{p_1} (0) ] \rangle
= {1 \over N_c} \langle \Tr [ W_{p_2} (0) ] \rangle ~,
\label{ZW}
\ee
where $N_c$ is the number of colours.

In a perfectly analogous way, one can also derive the high--energy scattering
amplitude for an elastic process involving two partons, which can be quarks,
antiquarks or gluons \cite{Nachtmann97,Meggiolaro01}.
For an antiquark, one simply has to substitute the Wilson line $W_p(b)$
with its complex conjugate $W^*_p(b)$: this is due to the fact that the
scattering amplitude of an antiquark in the external gluon field $A_\mu$
is equal to the scattering amplitude of a quark in the charge--conjugated
(C--transformed) gluon field $A'_\mu = -A^t_\mu = -A^*_\mu$.
In other words, going from quarks to antiquarks
corresponds just to the change from the fundamental representation $T_a$ of
$SU(N_c)$ to the complex conjugate representation $T'_a = -T^*_a$.
In the same way, going from quarks to gluons
corresponds just to the change from the fundamental representation $T_a$ of
$SU(N_c)$ to the adjoint representation $T^{(adj)}_a$. So, if the parton
is a gluon, one must substitute $W_p(b)$, the Wilson string in the
fundamental representation, with ${\cal V}_k (b)$, the Wilson string
in the adjoint representation [and the renormalization constant $Z_W$
with $Z_{\cal V} = \langle \Tr [{\cal V}_k (0)] \rangle / (N_c^2 - 1)$].

In what follows, to be definite, we shall consider the case of the quark--quark
scattering and we shall deal with the quantity
\be
g_{M (ij,kl)} (s; ~t) \equiv {1 \over Z_W^2}
\displaystyle\int d^2 \vec{z}_\perp e^{i \vec{q}_\perp \cdot \vec{z}_\perp}
\langle [ W_{p_1} (z_t) - {\bf 1} ]_{ij} [ W_{p_2} (0) - {\bf 1} ]_{kl} \rangle
~,
\label{gM}
\ee
in terms of which the scattering amplitude can be written as
\be
T_{fi} = \langle \psi_{i\alpha}(p'_1) \psi_{k\gamma}(p'_2) | \hat{T} |
\psi_{j\beta}(p_1) \psi_{l\delta}(p_2) \rangle
\mathop{\sim}_{s \to \infty}
-i \cdot 2s \cdot \delta_{\alpha\beta} \delta_{\gamma\delta}
\cdot g_{M (ij,kl)} (s; ~t) ~.
\label{scatt2}
\ee
At first sight, it could appear that the above expression (\ref{gM}) of the
quantity $g_M$ is essentially independent on the center--of--mass energy of
the two quarks and that the $s$ dependence of the scattering amplitude
is all contained in the kinematical factor $2s$ in front of the
integral in Eq. (\ref{scatt1}). This is clearly in contradiction with the
well--known fact that amplitudes in QCD have a very non--trivial $s$
dependence, whose origin lies in the infrared (IR) divergences typical
of $3 + 1$ dimensional gauge theories.
In more standard perturbative approaches to high--energy QCD, based on the
direct computation of Feynman diagrams in the high--energy limit, these IR
divergences are taken care of by restricting the rapidities of the intermediate
gluons to lie in between those of the two fast quarks (see, e.g.,
\cite{Cheng-Wu-book,Lipatov}).
The classical trajectories of two quarks with a non--zero mass $m$ and a
center--of--mass energy squared $s = 4 E^2$ are related by a finite Lorentz
boost with rapidity parameter $\log (s/m^2)$, so that the size of the rapidity
space for each intermediate gluon grows as $\log s$
and each Feynman diagram acquires an overall
factor proportional to some power of $\log s$, depending on the number of
intermediate gluon propagators.

In the case of the quantity (\ref{gM}), as was first pointed out by Verlinde
and Verlinde in \cite{Verlinde}, the IR singularity is originated by the
fact that the trajectories of the Wilson lines were taken to be lightlike and
therefore have an infinite distance in rapidity space.
One can regularize this infrared problem by giving the Wilson lines a small
timelike component, such that they coincide with the classical trajectories
for quarks with a non--zero mass $m$ (this is equivalent to consider two Wilson
lines forming a certain {\it finite} hyperbolic angle $\chi$ in Minkowski
space--time; of course, $\chi \to \infty$ when $s \to \infty$),
and, in addition, by letting them end after
some finite proper time $\pm T$ (and eventually letting $T \to \infty$).
Such a regularization of the IR singularities gives rise to an $s$ dependence
of the quantity $g_M$ defined in (\ref{gM}) and, therefore, to a non--trivial
$s$ dependence of the amplitude (\ref{scatt1}), as obtained by ordinary
perturbation theory \cite{Cheng-Wu-book,Lipatov} and as confirmed by the
experiments on hadron--hadron scattering processes. We refer the reader to
Refs.  \cite{Verlinde} and \cite{Meggiolaro97,Meggiolaro98}
for a detailed discussion about this point.

The direct evaluation of the expectation value (\ref{gM}) is a highly
non--trivial matter and it is also strictly connected with the renormalization
properties of Wilson--line operators \cite{Arefeva80,Korchemsky}.
A non--perturbative approach for the calculation of (\ref{gM}) has
been proposed and developed in Refs. \cite{Dosch,Berger},
in the framework of the so--called ``stochastic vacuum model''.
In two previous papers \cite{Meggiolaro97,Meggiolaro98} we proposed a new
approach, which consists in analytically continuing the scattering amplitude
from the Minkowskian to the Euclidean world, so opening the possibility of
studying the scattering amplitude non perturbatively
by well--known and well--established techniques available in the Euclidean
theory (e.g., by means of the formulation of the theory on the lattice).
This approach has been recently
adopted in Refs.~\cite{JP1,JP2}, in order to study the high--energy
scattering in strongly coupled gauge theories using the AdS/CFT correspondence,
and also in Ref.~\cite{Shuryak-Zahed}, in order to investigate
instanton--induced effects in QCD high--energy scattering.

More explicitly, in Refs. \cite{Meggiolaro97,Meggiolaro98} we have given
arguments showing that the expectation value of two {\it infinite} Wilson
lines, forming a certain hyperbolic angle $\chi$ in Minkowski space--time,
and the expectation value of two {\it infinite} Euclidean Wilson lines,
forming a certain angle $\theta$ in Euclidean four--space, are connected
by an analytic continuation in the angular variables.
This relation of analytic continuation was proven in
Ref. \cite{Meggiolaro97} for an Abelian gauge theory (QED) in the so--called
{\it quenched} approximation and for a non--Abelian gauge theory (QCD) up to
the fourth order in the renormalized coupling constant in perturbation theory;
a general proof was finally given in Ref. \cite{Meggiolaro98}.
The relation of analytic continuation between the amplitudes $g_M (\chi;~t)$
and $g_E (\theta;~t)$, in the Minkowski and the Euclidean world, was derived
in Refs. \cite{Meggiolaro97,Meggiolaro98} using {\it infinite} Wilson lines,
i.e., directly in the limit $T \to \infty$ and assuming that the amplitudes
were independent on $T$. In other words, the results derived in Refs.
\cite{Meggiolaro97,Meggiolaro98} apply to the cutoff--independent part of the
amplitudes.

On the contrary, in this paper we shall consider IR--regularized amplitudes
at any $T$ (including also possible divergent pieces when $T \to \infty$).
Generalizing the results of Ref. \cite{Meggiolaro98}, in Sect. 2 of this paper
we shall give the general proof that the expectation value of two
IR--regularized  Wilson lines, forming a certain hyperbolic angle in Minkowski
space--time, and the expectation value of two IR--regularized Euclidean Wilson
lines, forming a certain angle in Euclidean four--space, are connected by an
analytic continuation in the angular variables and in the IR cutoff $T$.
This result can be used to evaluate the IR--regularized high--energy scattering
amplitude directly in the Euclidean theory.
The conclusions and an outlook are given in Sect. 3.

\vfill\eject

\newsection{From Minkowskian to Euclidean theory}

\noindent
Let us consider the following quantity, defined in
Minkowski space--time:
\ba
g_M (p_1, p_2;~T;~t) &=& {M (p_1, p_2;~T;~t) \over Z_M(p_1;~T) Z_M(p_2;~T)} ~,
\nonumber \\
M (p_1, p_2;~T;~t) &=&
\displaystyle\int d^2 \vec{z}_\perp e^{i \vec{q}_\perp \cdot \vec{z}_\perp}
\langle [ W^{(T)}_{p_1} (z_t) - {\bf 1} ]_{ij}
[ W^{(T)}_{p_2} (0) - {\bf 1} ]_{kl} \rangle ~,
\label{M}
\ea
where $z_t = (0,0,\vec{z}_\perp)$ and $q = (0,0,\vec{q}_\perp)$, so that
$t = -\vec{q}_\perp^2 = q^2$. The Minkowskian four--momenta $p_1$ and $p_2$
are arbitrary four--vectors lying in the longitudinal plane $(x^0,x^1)$
[so that $\vec{p}_{1\perp} = \vec{p}_{2\perp} = \vec{0}_\perp$]
and define the trajectories of the two IR--regularized Wilson lines
$W^{(T)}_{p_1}$ and $W^{(T)}_{p_2}$:
\ba
W^{(T)}_{p_1} (z_t) &\equiv&
{\cal T} \exp \left[ -ig \displaystyle\int_{-T}^{+T}
A_\mu (z_t + {p_1 \over m} \tau) {p_1^\mu \over m} d\tau \right] ~;
\nonumber \\
W^{(T)}_{p_2} (0) &\equiv&
{\cal T} \exp \left[ -ig \displaystyle\int_{-T}^{+T}
A_\mu ({p_2 \over m} \tau) {p_2^\mu \over m} d\tau \right] ~.
\label{Wilson1}
\ea
$A_\mu = A_\mu^a T^a$ and $m$ is the quark {\it pole} mass.
$T$ is our IR cutoff. \\
$Z_M(p;~T)$ in Eq. (\ref{M}) is defined as ($N_c$ being the number of colours)
\be
Z_M(p;~T) \equiv {1 \over N_c} \langle \Tr [ W^{(T)}_{p} (z_t) ] \rangle
= {1 \over N_c} \langle \Tr [ W^{(T)}_{p} (0) ] \rangle ~.
\label{ZM}
\ee
(The last equality comes from the space--time translation invariance.)
This is a sort of Wilson--line's renormalization constant:
as shown in Ref. \cite{Meggiolaro01}, $Z_M(p~;T \to \infty)$ is the residue at
the pole (i.e., for $p^2 \to m^2$) of the unrenormalized quark propagator,
in the eikonal approximation.

In an analogous way, we can consider the following quantity, defined
in Euclidean four--space:
\ba
g_E (p_{1E}, p_{2E};~T;~t) &=& {E (p_{1E}, p_{2E};~T;~t) \over
Z_E(p_{1E};~T) Z_E(p_{2E};~T)} ~, \nonumber \\
E (p_{1E}, p_{2E};~T;~t) &=&
\displaystyle\int d^2 \vec{z}_\perp e^{i \vec{q}_\perp \cdot \vec{z}_\perp}
\langle [ \tilde{W}^{(T)}_{p_{1E}} (z_{t E}) - {\bf 1} ]_{ij}
[ \tilde{W}^{(T)}_{p_{2E}} (0) - {\bf 1} ]_{kl} \rangle_E ~,
\label{E}
\ea
where $z_{t E} = (0, \vec{z}_\perp, 0)$ and
$q_E = (0, \vec{q}_\perp, 0)$, so that: $t = -\vec{q}^2_\perp = -q_E^2$.
The expectation value $\langle \ldots \rangle_E$ must be intended now as a
functional integration with respect to the gauge variable $A^{(E)}_\mu =
A^{(E)a}_\mu T^a$ in the Euclidean theory.
The Euclidean four--momenta $p_{1E}$ and $p_{2E}$ are arbitrary four--vectors
lying in the plane $(x_1,x_4)$ [so that $\vec{p}_{1E\perp} = \vec{p}_{2E\perp}
= \vec{0}_\perp$] and define the trajectories of the two IR--regularized
Euclidean Wilson lines $\tilde{W}^{(T)}_{p_{1E}}$ and
$\tilde{W}^{(T)}_{p_{2E}}$:
\ba
\tilde{W}^{(T)}_{p_{1E}} (z_{t E}) &\equiv&
{\cal T} \exp \left[ -ig \displaystyle\int_{-T}^{+T}
A^{(E)}_{ \mu} (z_{t E} + {p_{1E} \over m} \tau) {p_{1E \mu} \over m} d\tau
\right] ~; \nonumber \\
\tilde{W}^{(T)}_{p_{2E}} (0) &\equiv&
{\cal T} \exp \left[ -ig \displaystyle\int_{-T}^{+T}
A^{(E)}_{ \mu} ({p_{2E} \over m} \tau) {p_{2E \mu} \over m} d\tau \right] ~.
\label{Wilson2}
\ea
$Z_E(p_E;~T)$ in Eq. (\ref{E}) is defined analogously to $Z_M(p;~T)$
in Eq. (\ref{ZM}):
\be
Z_E(p_E;~T) \equiv {1 \over N_c} \langle \Tr [ \tilde{W}^{(T)}_{p_E}
(z_{t E}) ] \rangle_E
= {1 \over N_c} \langle \Tr [ \tilde{W}^{(T)}_{p_E} (0) ] \rangle_E ~.
\label{ZE}
\ee
(The last equality comes from the translation invariance in Euclidean
four--space.)

We can now use the definition of the time--ordered exponential
in Eq. (\ref{Wilson1}) to explicitly write the Wilson lines
$W^{(T)}_{p_1}$ and $W^{(T)}_{p_2}$ as power series in the exponents
$g \cdot A$. Therefore, the quantity $M(p_1, p_2;~T;~t)$ is
defined to be the series $M = \sum_{n=1}^\infty \sum_{r=1}^\infty M_{(n,r)}$,
where $M_{(n,r)}$ is the contribution from the piece with $(g \cdot A)^n$
in the expansion of $W^{(T)}_{p_1}$ and from the piece with $(g \cdot A)^r$
in the expansion of $W^{(T)}_{p_2}$; it is given by:
\ba
\lefteqn{
M_{(n,r)} (p_1, p_2;~T;~t) =
(-ig)^{(n+r)} (T^{a_n} \ldots T^{a_1})_{ij}
(T^{b_r} \ldots T^{b_1})_{kl}
\displaystyle\int d^2 \vec{z}_\perp e^{i \vec{q}_\perp \cdot \vec{z}_\perp}
} \nonumber \\
& & \times \displaystyle\int_{-T}^{+T} d\tau_1 {p_1^{\mu_1} \over m} \ldots
\displaystyle\int_{-T}^{+T} d\tau_n {p_1^{\mu_n} \over m}
\displaystyle\int_{-T}^{+T} d\omega_1 {p_2^{\nu_1} \over m} \ldots
\displaystyle\int_{-T}^{+T} d\omega_r {p_2^{\nu_r} \over m}
\nonumber \\
& & \times \theta (\tau_n - \tau_{n-1}) \ldots \theta (\tau_2 - \tau_1)
\theta (\omega_r - \omega_{r-1}) \ldots \theta (\omega_2 - \omega_1)
\nonumber \\
& & \times \langle A_{\mu_n}^{a_n} (z_t + {p_1 \over m} \tau_n) \ldots
A_{\mu_1}^{a_1} (z_t + {p_1 \over m} \tau_1)
A_{\nu_r}^{b_r} ({p_2 \over m} \omega_r) \ldots
A_{\nu_1}^{b_1} ({p_2 \over m} \omega_1) \rangle ~.
\label{Mnr}
\ea
Analogously, the Euclidean quantity $E(p_{1E}, p_{2E};~T;~t)$ is defined to
be the series $E = \sum_{n=1}^\infty \sum_{r=1}^\infty E_{(n,r)}$, where
$E_{(n,r)}$ is the contribution from the pieces with
$(g \cdot A_{(E)})^n$ and $(g \cdot A_{(E)})^r$ in the expansions of the
Euclidean Wilson lines $\tilde{W}^{(T)}_{p_{1E}}$ and
$\tilde{W}^{(T)}_{p_{2E}}$ respectively; it is given by:
\ba
\lefteqn{
E_{(n,r)} (p_{1E}, p_{2E};~T;~t) =
(-ig)^{(n+r)} (T^{a_n} \ldots T^{a_1})_{ij}
(T^{b_r} \ldots T^{b_1})_{kl}
\displaystyle\int d^2 \vec{z}_\perp e^{i \vec{q}_\perp \cdot \vec{z}_\perp}
} \nonumber \\
& & \times \displaystyle\int_{-T}^{+T} d\tau_1 {p_{1E}^{\mu_1} \over m} \ldots
\displaystyle\int_{-T}^{+T} d\tau_n {p_{1E}^{\mu_n} \over m}
\displaystyle\int_{-T}^{+T} d\omega_1 {p_{2E}^{\nu_1} \over m} \ldots
\displaystyle\int_{-T}^{+T} d\omega_r {p_{2E}^{\nu_r} \over m}
\theta (\tau_n - \tau_{n-1}) \ldots \theta (\tau_2 - \tau_1)
\nonumber \\
& & \times \theta (\omega_r - \omega_{r-1}) \ldots \theta (\omega_2 - \omega_1)
\langle A_{(E) \mu_n}^{a_n} (z_{t E} + {p_{1E} \over m} \tau_n) \ldots
A_{(E) \mu_1}^{a_1} (z_{t E} + {p_{1E} \over m} \tau_1)
\nonumber \\
& & \times A_{(E) \nu_r}^{b_r} ({p_{2 E} \over m} \omega_r) \ldots
A_{(E) \nu_1}^{b_1} ({p_{2 E} \over m} \omega_1) \rangle_E ~.
\label{Enr}
\ea
It is known that, making use of the correspondence
\ba
A_0 (x) \rightarrow i A^{(E)}_{ 4}  (x_E) ~~ &,& ~~
A_k (x) \rightarrow A^{(E)}_{ k} (x_E) ~~~ \nonumber \\
{\rm with:}~~x^0 \rightarrow -i x_{E 4} ~~ &,& ~~
\vec{x} \rightarrow \vec{x}_E ~,
\label{correspondence}
\ea
between the Minkowski and the Euclidean world, the following relationship
is derived between the gluonic Green functions in the two theories:
\ba
\lefteqn{
\tilde{B}_{(1)}^{\mu_1} \ldots \tilde{B}_{(N)}^{\mu_N}
\langle A_{\mu_1}^{a_1} (\tilde{x}_{(1)}) \ldots
A_{\mu_N}^{a_N} (\tilde{x}_{(N)}) \rangle = } \nonumber \\
& & = B_{(1) E \mu_1} \ldots B_{(N) E \mu_N}
\langle A_{(E) \mu_1}^{a_1} (x_{(1) E}) \ldots
A_{(E) \mu_N}^{a_N} (x_{(N) E}) \rangle_E ~,
\label{Green0}
\ea
where $x_{(k) E} = (\vec{x}_{(k) E}, x_{(k) E 4})$ are Euclidean
four--coordinates and $B_{(k) E} = (\vec{B}_{(k) E}, B_{(k) E 4})$ are any
Euclidean four--vectors, while $\tilde{x}_{(k)}$ and $\tilde{B}_{(k)}$ are
Minkowski four--vectors defined as
\ba
\tilde{x}_{(k)} = (\tilde{x}_{(k)}^0, \vec{\tilde{x}}_{(k)}) &=&
(-i x_{(k) E 4}, \vec{x}_{(k) E}) ~, \nonumber \\
\tilde{B}_{(k)} = (\tilde{B}_{(k)}^0, \vec{\tilde{B}}_{(k)}) &=&
(-i B_{(k) E 4}, \vec{B}_{(k) E}) ~.
\label{4vect}
\ea
In our specific case, we can use Eq. (\ref{Green0}) to state that
\ba
\lefteqn{
{\tilde{p}_1^{\mu_1} \over m} \ldots {\tilde{p}_1^{\mu_n} \over m}
{\tilde{p}_2^{\nu_1} \over m} \ldots {\tilde{p}_2^{\nu_r} \over m}
\langle A_{\mu_n}^{a_n} (z_t + {\tilde{p}_1 \over m} \tau_n)
\ldots A_{\mu_1}^{a_1} (z_t + {\tilde{p}_1 \over m} \tau_1)
A_{\nu_r}^{b_r} ({\tilde{p}_2 \over m} \omega_r) \ldots
A_{\nu_1}^{b_1} ({\tilde{p}_2 \over m} \omega_1) \rangle }
\nonumber \\
& & = {p_{1E}^{\mu_1} \over m} \ldots {p_{1E}^{\mu_n} \over m}
{p_{2E}^{\nu_1} \over m} \ldots {p_{2E}^{\nu_r} \over m}
\langle A_{(E) \mu_n}^{a_n} (z_{t E} + {p_{1E} \over m} \tau_n)
\ldots A_{(E) \mu_1}^{a_1} (z_{t E} + {p_{1E} \over m} \tau_1)
\nonumber \\
& & \times A_{(E) \nu_r}^{b_r} ({p_{2 E} \over m} \omega_r) \ldots
A_{(E) \nu_1}^{b_1} ({p_{2 E} \over m} \omega_1) \rangle_E ~,
\label{Green1}
\ea
where $p_{kE} = (\vec{p}_{kE}, p_{kE4})$, for $k = 1,2$, are the two Euclidean
four--vectors introduced above and $\tilde{p}_k$ are the two corresponding
Minkowskian four--vectors, obtained according to Eq. (\ref{4vect}):
\be
\tilde{p}_k = (\tilde{p}_k^0, \vec{\tilde{p}}_k) =
(-i p_{kE4}, \vec{p}_{kE}) ~,~~~ {\rm for} ~k=1,2 ~.
\label{ptilde}
\ee
By virtue of the definitions (\ref{Mnr}) and (\ref{Enr}) for $M_{(n,r)}$
and $E_{(n,r)}$ respectively, Eq. (\ref{Green1}) implies that:
\be
E_{(n,r)} (p_{1E}, p_{2E};~T;~t) =
M_{(n,r)} (\tilde{p}_1, \tilde{p}_2;~T;~t) ~.
\label{EMnr}
\ee
This relation is valid for every couple of integer numbers $(n,r)$, so that,
more generally, $\sum_{n=1}^\infty \sum_{r=1}^\infty E_{(n,r)}(p_{1E},
p_{2E};~T;~t)$ $=$ $\sum_{n=1}^\infty \sum_{r=1}^\infty M_{(n,r)} (\tilde{p}_1,
\tilde{p}_2;~T;~t)$; and therefore, by definition:
\be
E (p_{1E}, p_{2E};~T;~t) =
M (\tilde{p}_1, \tilde{p}_2;~T;~t) ~.
\label{EMres1}
\ee
Moreover, one has that, changing the integration variable in the exponent of
the Wilson line from the {\it real} proper time $\tau$ to the {\it imaginary}
proper time $\tau' \equiv -i\tau$:
\ba
\lefteqn{
W^{(T)}_{\tilde{p}_1} (z_t)
= {\cal T} \exp \left[ -ig \displaystyle\int_{-T}^{+T}
A_\mu (z_t + {\tilde{p}_1 \over m} \tau) {\tilde{p}_1^\mu \over m}
d\tau \right] }
\nonumber \\
& &
= {\cal T} \exp \left[ -ig \displaystyle\int_{iT}^{-iT}
A_\mu (z_t + {\bar{p}_1 \over m} \tau') {\bar{p}_1^\mu \over m}
d\tau' \right]
\nonumber \\
& & = W^{(-iT)}_{\bar{p}_1} (z_t) ~,
\label{Wilson3}
\ea
and, similarly:
\be
W^{(T)}_{\tilde{p}_2} (0) = W^{(-iT)}_{\bar{p}_2} (0) ~,
\label{Wilson4}
\ee
where the Minkowskian four--vectors $\bar{p}_k$ are defined as:
\be
\bar{p}_k = i\tilde{p}_k = (p_{kE4}, i\vec{p}_{kE}) ~,~~~ {\rm for} ~k=1,2 ~.
\label{pbar}
\ee
The following prescription for the ${\cal T}$--ordered product of a bosonic
field $B$ in the imaginary domain is used:
\ba
{\cal T} B(\tau'_1) B(\tau'_2) &=& B(\tau'_1) B(\tau'_2) ~,
~~~{\rm if}~i\tau'_1 > i\tau'_2 ~; \nonumber \\
{\cal T} B(\tau'_1) B(\tau'_2) &=& B(\tau'_2) B(\tau'_1) ~,
~~~{\rm if}~i\tau'_1 < i\tau'_2 ~.
\label{Tord}
\ea
In other words, $\theta (\tau' = -i \tau) \equiv \theta (\tau)$, for every real
$\tau$: this prescription is used in order to keep the ${\cal T}$--ordering
unchanged when going from Minkowskian to Euclidean theory, $(x^0,\vec{x})
\rightarrow (-ix_{E4},\vec{x}_E)$. From the definition of
$M(p_1, p_2;~T;~t)$ given in Eq. (\ref{M}), one gets that:
\be
M(\tilde{p}_1, \tilde{p}_2;~T;~t)
= M(\bar{p}_1, \bar{p}_2;~-iT;~t) ~.
\label{Mres}
\ee
And therefore, from the Eq. (\ref{EMres1}) derived above:
\be
E (p_{1E}, p_{2E};~T;~t) =
M(\bar{p}_1, \bar{p}_2;~-iT;~t) ~.
\label{EMres2}
\ee
We also observe that, rescaling the four--momentum $p$ in the Wilson line
by a positive constant $\alpha$:
\ba
\lefteqn{
W^{(T)}_{\alpha p} (z_t)
= {\cal T} \exp \left[ -ig \displaystyle\int_{-T}^{+T}
A_\mu (z_t + {\alpha p \over m} \tau) {\alpha p^\mu \over m}
d\tau \right] }
\nonumber \\
& &
= {\cal T} \exp \left[ -ig \displaystyle\int_{-\alpha T}^{+\alpha T}
A_\mu (z_t + {p \over m} \tilde{\tau}) {p^\mu \over m}
d\tilde{\tau} \right]
\nonumber \\
& & = W^{(\alpha T)}_{p} (z_t) ~,~~~ \forall \alpha > 0 ~.
\label{Wilson5}
\ea
And similarly, for the Euclidean Wilson lines:
\be
\tilde{W}^{(T)}_{\alpha p_E} (z_{tE})
= \tilde{W}^{(\alpha T)}_{p_E} (z_{tE}) ~,~~~ \forall \alpha > 0 ~.
\label{Wilson6}
\ee
Of course $M$, considered as a general function of $p_1$, $p_2$ [and
$q = (0,0,\vec{q}_\perp)$], can only depend on the scalar quantities
constructed with the vectors $p_1$, $p_2$ and $q = (0,0,\vec{q}_\perp)$: the
only possibilities are $q^2 = -\vec{q}_\perp^2 = t$, $p_1 \cdot p_2$, $p_1^2$
and $p_2^2$, since $p_1 \cdot q = p_2 \cdot q = 0$. Therefore, from the result
(\ref{Wilson5}) found above, $M$ is forced to have the following form:
\be
M (p_1, p_2;~T;~t) = f_M \left( u_1^2 T^2, u_2^2 T^2,
(u_1 \cdot u_2) T^2;~t \right) ~,
\label{fM}
\ee
where $u_1 \equiv p_1 / m$ and $u_2 \equiv p_2 / m$.
[One can easily derive this by first introducing two different IR cutoffs,
$T_1$ and $T_2$, for the two Wilson lines $W_{p_1}$ and $W_{p_2}$, then by
using the result (\ref{Wilson5}) found above, so deriving the relation
$M (p_1, p_2;~T_1;~T_2;~t) =$ $ f_M \left(u_1^2 T_1^2, u_2^2 T_2^2,
(u_1 \cdot u_2) T_1 T_2;~t \right)$, and finally by putting
$T_1 = T_2 \equiv T$.] \\
For analogous reasons, $E$ must be of the form:
\be
E (p_{1E}, p_{2E};~T;~t)
= f_E \left(u_{1E}^2 T^2, u_{2E}^2 T^2,
(u_{1E} \cdot u_{2E}) T^2;~t \right) ~,
\label{fE}
\ee
where $u_{1E} \equiv p_{1E} / m$ and $u_{2E} \equiv p_{2E} / m$.
[A short remark about the notation: we have denoted everywhere the
scalar product by a ``$\cdot$'', both in the Minkowski and the Euclidean
world. Of course, when $A$ and $B$ are Minkowski four--vectors, then
$A \cdot B = A^\mu B_\mu = A^0 B^0 - \vec{A} \cdot \vec{B}$; while, if
$A_E$ and $B_E$ are Euclidean four--vectors, then
$A_E \cdot B_E = A_{E \mu} B_{E \mu} = \vec{A}_E \cdot \vec{B}_E +
A_{E 4} B_{E 4}$.]
Therefore, the relations (\ref{EMres1}) and (\ref{EMres2}) found above can be
re--formulated as follows [observing that $\tilde{p}^2 = -\bar{p}^2 =
-p_{E}^2$ and $\tilde{p}_1 \cdot \tilde{p}_2 = -\bar{p}_1 \cdot \bar{p}_2 =
-p_{1E} \cdot p_{2E}$]:
\be
f_E \left( u_{1E}^2 T^2, u_{2E}^2 T^2, (u_{1E} \cdot u_{2E}) T^2;~t \right)
= f_M \left( -u_{1E}^2 T^2, -u_{2E}^2 T^2,
-(u_{1E} \cdot u_{2E}) T^2;~t \right) ~.
\label{EMres3}
\ee
Since we finally want to obtain the expression (\ref{scatt1}) of the
scattering amplitude in the c.m.s. of the two quarks, taking their
initial trajectories along the $x^1$--axis, we {\it choose} $p_1$ and $p_2$
to be the four--momenta of the two particles with mass $m$, moving with speed
$\beta$ and $-\beta$ along the $x^1$--direction, i.e.,
\ba
p_1 &=& E (1,\beta,0,0) ~, \nonumber \\
p_2 &=& E (1,-\beta,0,0) ~,
\label{p12}
\ea
where $E = m / \sqrt{1 - \beta^2}$ (in units with $c=1$) is the
energy of each particle (so that: $s = 4E^2$).
We now introduce the hyperbolic angle $\psi$ [in the plane $(x^0,x^1)$]
of the trajectory of $W^{(T)}_{p_1}$: it is given by $\beta = \tanh \psi$.
We can give the explicit form of the Minkowski four--vectors
$u_1 = p_1/m$ and $u_2 = p_2/m$ in terms of the hyperbolic angle $\psi$:
\ba
u_1 = {p_1 \over m} &=& (\cosh \psi,\sinh \psi,0,0) ~, \nonumber \\
u_2 = {p_2 \over m} &=& (\cosh \psi,-\sinh \psi,0,0) ~.
\label{u12}
\ea
Clearly, $u_1^2 = u_2^2 = 1$ and
\be
u_1 \cdot u_2 = \cosh (2\psi) = \cosh \chi ~,
\label{chi}
\ee
where $\chi = 2\psi$ is the hyperbolic angle [in the plane $(x^0,x^1)$]
between the two trajectories of $W^{(T)}_{p_1}$ and $W^{(T)}_{p_2}$.

Analogously, in the Euclidean theory we {\it choose}
a reference frame in which the spatial vectors $\vec{p}_{1E}$ and
$\vec{p}_{2E} = -\vec{p}_{1E}$ are along the $x_1$--direction and,
moreover, $p_{1E}^2 = p_{2E}^2 = m^2$; that is,
\ba
p_{1E} &=& m (\sin \phi, 0, 0, \cos \phi ) ~; \nonumber \\
p_{2E} &=& m (-\sin \phi, 0, 0, \cos \phi ) ~,
\label{p12E}
\ea
where $\phi$ is the angle formed by each trajectory with the $x_4$--axis.
The value of $\phi$ is between $0$ and $\pi / 2$, so that the angle
$\theta = 2 \phi$ between the two Euclidean trajectories
$\tilde{W}^{(T)}_{p_{1E}}$ and $\tilde{W}^{(T)}_{p_{2E}}$
lies in the range $[0,\pi]$: it is always possible to make
such a choice by virtue of the $O(4)$ symmetry of the Euclidean theory.
The two four--momenta $u_{1E}$ and $u_{2E}$ are, therefore:
\ba
u_{1E} = {p_{1E} \over m} &=& (\sin \phi, 0, 0, \cos \phi ) ~; \nonumber \\
u_{2E} = {p_{2E} \over m} &=& (-\sin \phi, 0, 0, \cos \phi ) ~,
\label{u12E}
\ea
Clearly, $u_{1E}^2 = u_{2E}^2 = 1$ and
\be
u_{1E} \cdot u_{2E} = \cos \theta ~.
\label{theta}
\ee
With this choice, one has that:
\ba
\bar{p}_1 &=&
m (\cos {\theta \over 2},i\sin {\theta \over 2},0,0) =
m (\cosh {i\theta \over 2},\sinh {i\theta \over 2},0,0) ~; \nonumber \\
\bar{p}_2 &=&
m (\cos {\theta \over 2},-i\sin {\theta \over 2},0,0) =
m (\cosh {i\theta \over 2},-\sinh {i\theta \over 2},0,0) ~.
\label{pbar12}
\ea
A comparison with the expressions (\ref{u12}) for the Minkowski four--vectors
$u_1 = p_1/m$ and $u_2 = p_2/m$ reveals that $\bar{p}_1$ and $\bar{p}_2$ are
obtained from $p_1$ and $p_2$ after the following analytic continuation in the
angular variables is made:
\be
\chi \rightarrow i \theta ~.
\label{chitheta}
\ee
(We remind that $\phi = \theta/2$ and $\psi = \chi/2$.)
Therefore, if we denote with $M(\chi;~T;~t)$ the value of $M(p_1, p_2;~T;~t)$
for $p_1$ and $p_2$ given by Eq. (\ref{p12}) and we also denote with
$E(\theta;~T;~t)$ the value of $E (p_{1E}, p_{2E};~T;~t)$ for $p_{1E}$ and
$p_{2E}$ given by Eq. (\ref{p12E}), we find, using the result (\ref{EMres2})
derived above:
\be
E (\theta;~T;~t) = M (\chi \to i\theta;~T \to -iT;~t) ~.
\label{EMres4}
\ee
This is, of course, in agreement with the relation (\ref{EMres3}) found above,
observing that $M(\chi~;~T;~t) = f_M(T^2, T^2, T^2 \cosh \chi;~t)$ and
$E(\theta~;~T;~t) = f_E(T^2, T^2, T^2 \cos \theta;~t)$.

Let us consider, now, the Wilson--line's renormalization constant $Z_M(p;~T)$:
\be
Z_M(p;~T) \equiv {1 \over N_c} \langle \Tr [ W^{(T)}_{p} (0) ] \rangle ~.
\label{ZMbis}
\ee
Again, we can use the definition of the time--ordered exponential in Eq.
(\ref{Wilson1}) to expand the Wilson line $W^{(T)}_{p} (0)$ in powers of the
exponent $g \cdot A$. The quantity $Z_M(p;~T)$ is thus defined to be the series
$Z_M(p;~T) = \sum_{n=1}^{\infty} Z_M^{(n)}(p;~T)$, where $Z_M^{(n)}(p;~T)$
is the contribution from the piece with $(g \cdot A)^n$ in the expansion of
$W^{(T)}_{p} (0)$; it is given by:
\ba
\lefteqn{
Z_M^{(n)}(p;~T) =
{(-ig)^n \over N_c} \Tr (T^{a_n} \ldots T^{a_1})
\displaystyle\int_{-T}^{+T} d\tau_1 {p^{\mu_1} \over m} \ldots
\displaystyle\int_{-T}^{+T} d\tau_n {p^{\mu_n} \over m}
} \nonumber \\
& & \times \theta (\tau_n - \tau_{n-1}) \ldots \theta (\tau_2 - \tau_1)
\langle A_{\mu_n}^{a_n} ({p \over m} \tau_n) \ldots
A_{\mu_1}^{a_1} ({p \over m} \tau_1) \rangle ~.
\label{ZMn}
\ea
In the Euclidean theory we have, analogously:
\be
Z_E(p_E;~T) \equiv {1 \over N_c} \langle \Tr [ \tilde{W}^{(T)}_{p_{E}} (0) ]
\rangle_E ~,
\label{ZEbis}
\ee
and $Z_E(p_E;~T) = \sum_{n=1}^{\infty} Z_E^{(n)}(p_E;~T)$, with
\ba
\lefteqn{
Z_E^{(n)}(p_E;~T) =
{(-ig)^n \over N_c} \Tr (T^{a_n} \ldots T^{a_1})
\displaystyle\int_{-T}^{+T} d\tau_1 {p_E^{\mu_1} \over m} \ldots
\displaystyle\int_{-T}^{+T} d\tau_n {p_E^{\mu_n} \over m}
} \nonumber \\
& & \times \theta (\tau_n - \tau_{n-1}) \ldots \theta (\tau_2 - \tau_1)
\langle A_{(E) \mu_n}^{a_n} ({p_E \over m} \tau_n) \ldots
A_{(E) \mu_1}^{a_1} ({p_E \over m} \tau_1) \rangle_E ~.
\label{ZEn}
\ea
Using Eq. (\ref{Green0}), we can derive the following relation:
\ba
\lefteqn{
{\tilde{p}^{\mu_1} \over m} \ldots {\tilde{p}^{\mu_n} \over m}
\langle A_{\mu_n}^{a_n} ({\tilde{p} \over m} \tau_n) \ldots
A_{\mu_1}^{a_1} ({\tilde{p} \over m} \tau_1) \rangle
} \nonumber \\
& & = {p_E^{\mu_1} \over m} \ldots {p_E^{\mu_n} \over m}
\langle A_{(E) \mu_n}^{a_n} ({p_E \over m} \tau_n) \ldots
A_{(E) \mu_1}^{a_1} ({p_E \over m} \tau_1) \rangle_E ~,
\label{Green2}
\ea
where, as usual, $p_E = (\vec{p}_E, p_{E4})$ and
$\tilde{p} = (\tilde{p}^0, \vec{\tilde{p}})
= (-i p_{E4}, \vec{p}_E)$.
From this relation we obtain
\be
Z_E^{(n)} (p_E;~T) = Z_M^{(n)} (\tilde{p};~T) ~.
\label{ZEMn}
\ee
This relation is valid for every integer number $n$, so that we also have,
more generally, $\sum_{n=1}^{\infty} Z_E^{(n)} (p_E;~T)$ $=$
$\sum_{n=1}^{\infty} Z_M^{(n)} (\tilde{p};~T)$; and therefore, by definition:
\be
Z_E (p_E;~T) = Z_M (\tilde{p};~T) ~.
\label{ZEMres1}
\ee
Moreover, from Eq. (\ref{Wilson3}) one derives that
$Z_M (\tilde{p};~T) = Z_M (\bar{p};~-iT)$, where, as usual,
$\bar{p} = i\tilde{p}$. And therefore, from Eq. (\ref{ZEMres1}):
\be
Z_E (p_E;~T) = Z_M (\bar{p};~-iT) ~.
\label{ZEMres2}
\ee
From the definition (\ref{ZM}), $Z_M (p;~T)$, considered as a
function of a general four--vector $p$, is a scalar function constructed
with the only four--vector $u = p/m$. In addition, by virtue of the property
(\ref{Wilson5}) of the Wilson lines, one has that $Z_M (\alpha p;~T) =
Z_M (p;~\alpha T)$ for every positive $\alpha$. Therefore, $Z_M (p;~T)$
is forced to have the form
\be
Z_M (p;~T) = H_M (u^2 T^2) ~,
\label{HM}
\ee
where $u = p/m$.
In a perfectly analogous way, for the Euclidean case we have that:
\be
Z_E (p_E;~T) = H_E (u_E^2 T^2) ~,
\label{HE}
\ee
where $u_E = p_E/m$. Therefore, the relations (\ref{ZEMres1}) and
(\ref{ZEMres2}) found above can be re--formulated as follows
[observing that $\tilde{p}^2 = -\bar{p}^2 = -p_E^2$]:
\be
H_E (u_E^2 T^2) = H_M (-u_E^2 T^2) ~.
\label{ZEMres3}
\ee
Therefore, if we denote with  $Z_W(T)$ the value of $Z_M(p_1;~T)$ or
$Z_M(p_2;~T)$, for $p_1$ and $p_2$ given by Eq. (\ref{p12}), and we also denote
with $Z_{W E}(T)$ the value of $Z_E(p_{1E};~T)$ or $Z_E(p_{2E};~T)$,
for $p_{1E}$ and $p_{2E}$ given by Eq. (\ref{p12E}), i.e.,
\ba
Z_W(T) &\equiv& Z_M(p_1;~T) = Z_M(p_2;~T) = H_M(T^2) ~; \nonumber \\
Z_{W E}(T) &\equiv& Z_E(p_{1E};~T) = Z_E(p_{2E};~T) = H_E(T^2) ~,
\label{ZW-ZWE}
\ea
we find, using the result (\ref{ZEMres2}) derived above:
\be
Z_{W E}(T) = Z_W(-iT) ~.
\label{ZEMres4}
\ee
Combining this identity with Eq. (\ref{EMres4}), we find that the Minkowskian
and the Euclidean amplitudes, defined by Eqs. (\ref{M}), (\ref{E}) and
(\ref{ZW-ZWE}), with $p_1$ and $p_2$ given by Eq. (\ref{p12}) and $p_{1E}$
and $p_{2E}$ given by Eq. (\ref{p12E}), i.e.,
\be
g_M (\chi;~T;~t) \equiv {M (\chi;~T;~t) \over [Z_W(T)]^2} ~, ~~~
g_E (\theta;~T;~t) \equiv {E (\theta;~T;~t) \over [Z_{W E}(T)]^2} ~,
\label{gM-gE}
\ee
are connected by the following relation:
\ba
g_E (\theta;~T;~t) &=& g_M (\chi \to i\theta;~T \to -iT;~t) ~;
\nonumber \\
{\rm or:}~~
g_M (\chi;~T;~t) &=& g_E (\theta \to -i\chi;~T \to iT;~t) ~.
\label{final}
\ea
We have derived the relation (\ref{final}) of analytic continuation for a
non--Abelian gauge theory with gauge group $SU(N_c)$. It is clear, from the
derivation given above, that the same result is valid also for an Abelian
gauge theory (QED).

Moreover, even if the result (\ref{final}) has been explicitly derived for the
case of the quark--quark scattering, it is immediately generalized to the more
generale case of the parton--parton scattering, where each parton can be a
quark, an antiquark or a gluon. In fact, as explained in the Introduction,
one simply has to use a proper Wilson line for each parton: $W_p (b)$, the
Wilson string in the fundamental representation $T^a$, for a quark;
$W^*_p (b)$, the Wilson string in the complex conjugate representation
$T'_a = -T^*_a$, for an antiquark; and ${\cal V}_k (b)$, the Wilson string
in the adjoint representation $T^{(adj)}_a$, for a gluon.
The proof leading to Eq. (\ref{final}) is then repeated step by step, after
properly modifying the definitions (\ref{Wilson1}) of the Wilson lines.
[If the parton is a gluon, one must substitute the quark mass $m$ appearing
in all previous formulae with an arbitrarily small mass $\mu \to 0$. The
IR cutoff appears in all expressions in the form of the ratio $T/\mu$ for
a gluon and $T/m$ for a quark/antiquark.]

\vfill\eject

\newsection{Conclusions and outlook}

\noindent
In this paper we have completely generalized the results of Ref.
\cite{Meggiolaro98}, where we derived a relation of analytic continuation
between the amplitudes $g_M (\chi;~t)$ and $g_E (\theta;~t)$, in the
Minkowski and the Euclidean world, using {\it infinite} Wilson lines, i.e.,
directly in the limit $T \to \infty$ and assuming that the amplitudes
were independent on $T$. In other words, we can claim that the results
of Ref. \cite{Meggiolaro98} apply to the cutoff--independent part of the
amplitudes, while, in this paper, we have derived the relation (\ref{final})
of analytic continuation between IR--regularized amplitudes at any $T$.

The result (\ref{final}) found in this paper can be used to evaluate
the IR--regularized high--energy parton--parton scattering amplitude
directly in the Euclidean theory. In fact, the IR--regularized
high--energy scattering amplitude is given (e.g., for the case of the
quark--quark scattering) by
\be
T_{fi} = \langle \psi_{i\alpha}(p'_1) \psi_{k\gamma}(p'_2) | \hat{T} |
\psi_{j\beta}(p_1) \psi_{l\delta}(p_2) \rangle
\mathop{\sim}_{s \to \infty}
-i \cdot 2s \cdot \delta_{\alpha\beta} \delta_{\gamma\delta}
\cdot g_M (\chi \to \infty;~T \to \infty;~t) ~,
\label{scatt4}
\ee
where the quantity $g_M (\chi;~T;~t)$, defined by Eq. (\ref{M}), is
essentially a correlation function of two IR--regularized Wilson lines
forming a certain hyperbolic angle $\chi$ in Minkowski space--time.
For deriving the dependence on $s$ one exploits the fact that the hyperbolic
angle $\chi$ is a function of $s$. In fact, from $s = 4E^2$,
$E = m/\sqrt{1 - \beta^2}$, and $\beta = \tanh (\chi/2)$ [see Eqs. (\ref{p12}),
(\ref{u12}) and (\ref{chi})], one immediately finds that:
\be
s = 2 m^2 ( \cosh \chi + 1 ) ~.
\label{s-chi}
\ee
Therefore, in the high--energy limit $s \to \infty$ (or $\chi \to \infty$,
i.e., $\beta \to 1$), the hyperbolic angle $\chi$ is essentially equal
to the logarithm of $s/m^2$ (for a non--zero quark mass $m$):
\be
\chi \mathop{\sim}_{s \to \infty} \log \left( {s \over m^2} \right) ~.
\label{logs}
\ee
The quantity $g_M (\chi;~T;~t)$ is linked to the corresponding Euclidean
quantity $g_E (\theta;~T;~t)$, defined by Eq. (\ref{E}),
by the analytic continuation (\ref{final}) in the angular variables and in
the IR cutoff $T$.
Therefore, one can start by evaluating $g_E (\theta;~T;~t)$, which is
essentially a correlation function of two IR--regularized Wilson lines
forming a certain angle $\theta$ in Euclidean four--space, then by continuing
this quantity into Minkowski space--time by rotating the Euclidean angular
variable clockwise, $\theta \to -i \chi$, and the IR cutoff (Euclidean proper
time) anticlockwise, $T \to iT$: in such a way one reconstructs the Minkowskian
quantity $g_M (\chi;~T;~t)$.
As was pointed out in \cite{JP2}, one should note that {\it a priori} there is
an ambiguity in making such an analytical continuation, depending on the
precise choice of the path. This phenomenon does not
appear when the Euclidean correlation function $g_E(\theta;~T;t)$ has only
simple poles in the complex $\theta$--plane, but in some cases the analiticity
structure can contain branch cuts in the complex plane, which must be taken
into account: we refer the reader to Ref. \cite{JP2} for a full
discussion about this point.

We want to conclude by making a remark about the problem of the IR
divergences which appear in the high--energy scattering amplitudes.

A well--known feature of the parton--parton scattering amplitude is its
IR divergence, which, as we have already said in the Introduction, is typical
of $3 + 1$ dimensional gauge theories and which, in our formulation,
manifests itself in the IR singularity of the correlation
function of two Wilson lines  when $T \to \infty$.
In many cases these IR divergences can be factorized out.

As suggested in Ref. \cite{JP2}, an alternative way to eliminate this
cutoff dependence is to consider an IR--finite physical quantity, like
the scattering amplitude of two colourless states in gauge theories,
e.g., two $q \bar{q}$ meson states.
It was shown in Ref. \cite{Nachtmann97} that the high--energy meson--meson
scattering amplitude can be approximately reconstructed by first evaluating,
in the eikonal approximation, the scattering amplitude of two $q \bar{q}$
pairs, of given transverse sizes $\vec{R}_{1\perp}$ and $\vec{R}_{2\perp}$
respectively, and then folding this amplitude with two proper
wave functions $\omega_1 (\vec{R}_{1\perp})$ and $\omega_2 (\vec{R}_{2\perp})$
describing the two interacting mesons. It turns out that the high--energy
scattering amplitude of two $q \bar{q}$ pairs of transverse sizes
$\vec{R}_{1\perp}$ and $\vec{R}_{2\perp}$, and impact--parameter distance
$\vec{z}_\perp$, is governed by the correlation function of two Wilson loops
${\cal W}_1$ and ${\cal W}_2$, which follow the classical straight lines for
quark (antiquark) trajectories \cite{Nachtmann97,Dosch}:
\ba
{\cal W}_1 &\to& X_{\pm 1}^\mu(\tau) = z_t^\mu + {p_1^\mu \over m} \tau
\pm {R_{1t}^\mu \over 2} ~; \nonumber \\
{\cal W}_2 &\to& X_{\pm 2}^\mu(\tau) = {p_2^\mu \over m} \tau
\pm {R_{2t}^\mu \over 2} ~,
\label{loops}
\ea
[where $R_{1t} = (0,0,\vec{R}_{1\perp})$ and $R_{2t} = (0,0,\vec{R}_{2\perp})$]
and close at proper times $\tau = \pm T$. \\
The same analytical continuation (\ref{final}), that we have derived for the
case of Wilson lines, is, of course, expected to apply also
to the Wilson--loop correlators: the proof can be repeated going
essentially through the same steps as in the previous section, after
adapting the definitions (\ref{Wilson1}) from the case of Wilson lines
to the case of Wilson loops.
However, in this case the cutoff dependence on $T$ is expected to be removed
together with the related IR divergence which was present for the
case of Wilson lines.

In our opinion, the high--energy scattering
problem could be directly investigated on the lattice using this
Wilson--loop formulation.
A further advantage of the Wilson--loop formulation, which makes it suitable
to be studied on the lattice, is that, contrary to the Wilson--line
formulation, it is manifestly gauge--invariant.
(In the case of the parton--parton scattering amplitude, gauge invariance can
be restored, at least for the {\it diffractive}, i.e., no-colour-exchange,
part proportional to $\langle \Tr [W_{p_1}(z_t) - {\bf 1}] \Tr [W_{p_2}(0)
- {\bf 1}] \rangle$, by requiring that the gauge transformations at both ends
of the Wilson lines are the same \cite{Nachtmann91,Verlinde}.)
A considerable progress is expected along this line in the near future.

\bigskip
\noindent {\bf Acknowledgements}
\smallskip

I would like to thank R. Peschanski for useful discussions (during the
{\it ``Sixth Workshop on Non--Perturbative Quantum Chromodynamics''},
Paris, France, June 5th--9th, 2001), which have inspired this work.

\vfill\eject

{\renewcommand{\Large}{\normalsize}
}

\vfill\eject

\end{document}